\documentclass[conference]{IEEEtran}
\IEEEoverridecommandlockouts
\usepackage{amsmath,amssymb,amsfonts}
\usepackage{algorithmic}
\usepackage{graphicx}
\usepackage{textcomp}
\usepackage{xcolor}
\usepackage{multirow}
\usepackage{booktabs}
\usepackage{array}
\usepackage[normalem]{ulem}
\usepackage[utf8]{inputenc}
\usepackage{pifont}
\useunder{\uline}{\ul}{}

\title{Detecting Cybersecurity Threats by Integrating Explainable AI with SHAP Interpretability and Strategic Data Sampling}

\author{
\IEEEauthorblockN{\normalsize Norrakith Srisumrith}
\IEEEauthorblockA{\textit{\small Dept. of Digital Network and Information Security Management}\\
\textit{\small King Mongkut's University of Technology North Bangkok}\\
\small Bangkok, Thailand\\
\small norrakith.s@email.kmutnb.ac.th}
\and
\IEEEauthorblockN{\normalsize Sunantha Sodsee}
\IEEEauthorblockA{\textit{\small Dept. of Digital Network and Information Security Management}\\
\textit{\small King Mongkut's University of Technology North Bangkok}\\
\small Bangkok, Thailand\\
\small sunantha.s@itd.kmutnb.ac.th}
\thanks{
This manuscript has been accepted for publication in ICTIS 2026 
and will appear in Springer Lecture Notes in Networks and Systems (LNNS). 
The final authenticated version will be available via SpringerLink.
}
}
\begin{document}

\maketitle

\begin{abstract}
The critical need for transparent and trustworthy machine learning in cybersecurity operations drives the development of this integrated Explainable AI (XAI) framework. Our methodology addresses three fundamental challenges in deploying AI for threat detection: handling massive datasets through Strategic Sampling Methodology that preserves class distributions while enabling efficient model development; ensuring experimental rigor via Automated Data Leakage Prevention that systematically identifies and removes contaminated features; and providing operational transparency through Integrated XAI Implementation using SHAP analysis for model-agnostic interpretability across algorithms. Applied to the CIC-IDS2017 dataset, our approach maintains detection efficacy while reducing computational overhead and delivering actionable explanations for security analysts. The framework demonstrates that explainability, computational efficiency, and experimental integrity can be simultaneously achieved, providing a robust foundation for deploying trustworthy AI systems in security operations centers where decision transparency is paramount.
\end{abstract}


\section{Introduction}

The exponential growth of network-based attacks and sophisticated cyber threats has fundamentally transformed cybersecurity from a traditional perimeter defense challenge to a complex data-driven detection problem \cite{sommer2010outside,anderson2018ember}. With the increasing volume and sophistication of cyber attacks, machine learning approaches have demonstrated remarkable success in identifying malicious activities across diverse network environments \cite{buczak2016survey,ferrag2020deep}. However, the practical deployment of these advanced detection systems faces significant hurdles due to the inherent "black box" nature of complex models, particularly ensemble methods and deep learning architectures that achieve state-of-the-art performance \cite{rudin2019stop,adadi2018peeking}. This interpretability gap becomes critically problematic in security operations centers (SOCs), where analysts require not only high detection accuracy but also transparent, understandable decisions that they can validate, trust, and act upon during incident response scenarios \cite{arp2020dos,miller2019explanation}. The challenge is further compounded by the massive scale of modern network datasets, such as CIC-IDS2017 with over 2.8 million records, which necessitates efficient sampling strategies without compromising detection efficacy or introducing statistical biases \cite{distilxids2026,ma2025fsllm}.

Recent advancements in Explainable AI (XAI) have begun to address these transparency concerns, with SHAP (SHapley Additive exPlanations) emerging as a prominent framework for model interpretation due to its strong theoretical foundations and model-agnostic capabilities \cite{lundberg2017unified,lundberg2020local}. However, current implementations often treat explainability as a post-hoc analysis layer rather than integrating it throughout the machine learning pipeline \cite{gaitan2023explainable,uddin2024explainable}. Simultaneously, feature selection methodologies have evolved to not only improve computational efficiency but also enhance model interpretability by identifying the most discriminative characteristics while reducing dimensionality \cite{peng2005feature,li2017feature}. The work of Zhou et al. \cite{zhou2025adaptive} on adaptive ensemble learning and Ma et al. \cite{ma2025fsllm} on feature selection with large language models represents important steps toward practical deployment, yet these approaches often prioritize performance optimization over operational transparency. Furthermore, critical challenges around data leakage prevention, as highlighted by Pendlebury et al. \cite{pendlebury2019tesseract} and Arp et al. \cite{arp2020dos}, underscore the importance of rigorous experimental methodologies that ensure model evaluations reflect real-world deployment scenarios.

Our research addresses these challenges through an integrated Explainable AI framework that combines strategic sampling, state-of-the-art performance, and comprehensive interpretability. The main contributions are:

\begin{enumerate}
\item \textbf{Strategic Sampling Methodology}: A novel stratified sampling approach that maintains class distribution while enabling efficient model development
\item \textbf{Automated Data Leakage Prevention}: Comprehensive detection and removal of 37\% potentially leaky features.
\item \textbf{Systematic Algorithm Evaluation}: Multi-stage validation across XGBoost, Random Forest, and Logistic Regression
\item \textbf{Feature Selection Optimization}: Comparative analysis of MRMR and Chi2 methods across multiple reduction percentages
\item \textbf{Integrated XAI Implementation}: SHAP analysis for model-agnostic interpretability across all algorithms
\end{enumerate}

\section{Related Work}

\subsection{Machine Learning in Cybersecurity}

Machine learning (ML) methods for cybersecurity threat detection have developed considerably in the last decade from traditional signature-based approaches to advanced behavior based anomaly detection systems \cite{buczak2016survey,khraisat2019survey}. Early techniques in the detection models targeted traditional algorithms such as Decision Trees, Naive Bayes, and SVMs which had reasonable detection capabilities but suffered from adaptive attack patterns or scalability to large network scale \cite{sommer2010outside}.

The emergence of comprehensive benchmark datasets like CIC-IDS2017 \cite{sharafaldin2018toward, ring2019toolset} has enabled more rigorous evaluation of ML approaches in realistic network settings. Recent studies have demonstrated varying performance levels, with ensemble methods and deep learning architectures showing particular promise. Ferrag et al. \cite{ferrag2020deep} conducted an extensive comparative study showing that deep learning approaches could achieve up to 97.16\% accuracy on CIC-IDS2017, while Liu and Lang \cite{liu2019machine} provided a comprehensive survey highlighting the trade-offs between different ML paradigms in security contexts.

Recent advancements have focused on addressing the computational challenges of large-scale security datasets. DistilXIDS \cite{distilxids2026} introduced an efficient, lightweight transformer-based language model for real-time network intrusion detection, achieving high performance while maintaining explainability. Similarly, Ma et al. \cite{ma2025fsllm} proposed an IoT intrusion detection framework based on feature selection and large language models fine-tuning, achieving improved efficiency through strategic feature reduction. These approaches represent significant steps toward practical deployment of ML systems in operational security environments.

\subsection{Ensemble Methods and Hybrid Models for Intrusion Detection}

Ensemble learning has emerged as a powerful paradigm for intrusion detection, leveraging multiple models to improve accuracy and robustness beyond what individual classifiers can achieve. Recent work has explored various ensemble strategies, including voting, bagging, boosting, and stacking approaches. Among these, hybrid ensemble models that combine multiple ensemble techniques have shown particular promise in handling the complexity and variability of modern network threats.

Ahmed et al. \cite{hybridensemble2024} proposed a novel Hybrid Adaptive Ensemble for Intrusion Detection (HAEnID) that integrates three distinct ensemble methods: Stacking Ensemble (SEM), Bayesian Model Averaging (BMA), and Conditional Ensemble Method (CEM). Their approach demonstrated exceptional performance on the CIC-IDS2017 dataset, with BMA achieving 98.79\% accuracy when using 20 selected features. The HAEnID model incorporates an adaptive mechanism that allows ensemble components to evolve as network traffic patterns change, providing resilience against emerging attack vectors. Furthermore, the model integrates Explainable AI techniques including SHAP and LIME to enhance interpretability, addressing the "black box" problem common in complex ensemble systems.

A key innovation in the HAEnID framework is its strategic use of SMOTE (Synthetic Minority Oversampling Technique) \cite{chawla2002smote} to address class imbalance in multi-class classification tasks. By generating synthetic samples for minority attack classes, the model ensures balanced representation during training, significantly improving detection rates for rare attack types. This sampling strategy, combined with comprehensive feature selection methods including correlation analysis, information gain, and PCA, enables the model to maintain high performance while reducing computational complexity. The model represents a significant advancement in ensemble-based intrusion detection, demonstrating that hybrid approaches can achieve both high accuracy and practical interpretability.

\subsection{Explainable AI in Security Applications}

The growing complexity of machine learning models has spurred substantial research interest in explainable AI techniques, particularly in security-critical applications where decision transparency is essential \cite{rudin2019stop, adadi2018peeking}. The fundamental challenge lies in balancing model performance with interpretability, as noted by Miller \cite{miller2019explanation}, who emphasized that explanations must be comprehensible to human operators while accurately representing model reasoning.

SHAP (SHapley Additive exPlanations) has emerged as a prominent framework for model interpretation due to its strong theoretical foundations and model-agnostic capabilities \cite{lundberg2017unified, lundberg2020local}. Lundberg et al. \cite{lundberg2020local} demonstrated that TreeSHAP could provide consistent and accurate feature importance explanations for tree-based models, making it particularly suitable for cybersecurity applications where random forests and gradient boosting machines are commonly employed.

Recent research has produced several notable XAI implementations for intrusion detection. Mahmoud et al. \cite{mahmoud2025xi2s} proposed XI2S-IDS, a two-stage framework that separates binary anomaly detection from multi-class attack classification, achieving 99.81\% accuracy on CICIDS2017 while using SHAP to provide global explanations for the binary classifier. Their work specifically addresses the challenge of detecting low-frequency attacks through targeted multi-class training on attack-only data. Similarly, in the domain of hybrid ensemble models, recent work has demonstrated that combining multiple algorithms with SHAP-based explanations can yield both high performance and interpretability. This approach typically achieves superior detection rates while maintaining transparency through feature importance analysis and decision rationalization.

Gaitán-Cárdenas et al. \cite{gaitan2023explainable} developed explainable AI-based intrusion detection systems for cloud and IoT environments, demonstrating how XAI can enhance trust in automated security decisions. Uddin et al. \cite{uddin2024explainable} applied explainable transformer-based models for phishing email detection using a large language model approach, achieving high accuracy while providing natural language explanations for security alerts. These approaches highlight the growing recognition of XAI as essential for operational security systems.

\subsection{Feature Selection and Sampling Strategies}

Feature selection has long been recognized as a critical component of effective machine learning systems, particularly in high-dimensional domains like cybersecurity \cite{guyon2003introduction, li2017feature}. Traditional approaches include filter methods (such as Chi-squared and mutual information), wrapper methods, and embedded methods, each with distinct advantages and limitations.

Peng et al. \cite{peng2005feature} introduced the MRMR (Maximum Relevance Minimum Redundancy) criterion, which selects features that have high mutual information with the target while having low mutual information with other features. This approach has proven particularly effective in cybersecurity contexts where feature interdependencies can obscure detection patterns. Forman \cite{forman2003extensive} provided an extensive empirical study of feature selection metrics, demonstrating that the choice of selection method significantly impacts model performance and interpretability.

Recent work has explored the intersection of feature selection and interpretability. Ma et al. \cite{ma2025fsllm} developed an IoT intrusion detection framework combining feature selection with large language model fine-tuning, showing that careful feature selection can improve both performance and explainability. Their approach maintained high accuracy while providing clearer decision boundaries for security analysis. Similarly, DistilXIDS \cite{distilxids2026} incorporated efficient feature processing within their transformer architecture, demonstrating that strategic feature handling can enhance real-time detection capabilities.

Sampling strategies have gained increased attention as cybersecurity datasets continue to grow in scale. Ahmed et al. \cite{hybridensemble2024} employed SMOTE \cite{chawla2002smote} to address class imbalance in their hybrid ensemble model, demonstrating that synthetic oversampling of minority classes can significantly improve multi-class classification performance for rare attack types. The computational efficiency demonstrated in Chen et al.'s RIDE framework \cite{chen2023ride} highlights the importance of strategic data handling for real-time intrusion detection. Their hardware-accelerated approach preserved detection performance while enabling efficient inference, making real-time model deployment feasible in resource-constrained environments.

\subsection{Data Leakage Prevention in Cybersecurity ML}

Data leakage represents a significant challenge in cybersecurity machine learning, where subtle temporal dependencies and feature-target relationships can artificially inflate performance metrics \cite{arp2020dos}. Pendlebury et al. \cite{pendlebury2019tesseract} introduced TESSERACT, a framework for eliminating experimental bias in malware classification across space and time, highlighting the critical importance of proper dataset partitioning and evaluation methodologies.

Anderson and Roth \cite{anderson2018ember} emphasized the need for careful dataset construction in their work on EMBER, an open dataset for training static PE malware machine learning models. Their methodology includes rigorous procedures for preventing data leakage through temporal splitting and feature analysis, setting important precedents for the cybersecurity ML community.

Recent work by Arp et al. \cite{arp2020dos} provided comprehensive guidelines for machine learning in computer security, highlighting common pitfalls in experimental design and evaluation. Their "dos and don'ts" framework has become an essential reference for researchers working in this domain, emphasizing the importance of realistic evaluation scenarios and proper handling of temporal aspects in security data.

\subsection{Integration Gaps and Research Opportunities}

While recent research has advanced individual components of IDS pipelines, important gaps remain in their holistic, large-scale, and evaluation‑rigorous integration for enterprise‑grade deployment. Existing frameworks often optimize a single aspect (e.g., architecture, imbalance handling, or XAI) or combine techniques without explicitly addressing the computational and operational constraints of modern Security Operations Centers (SOCs). For example, HAEnID integrates SMOTE‑based class rebalancing, multi‑step feature selection, and SHAP/LIME explanations within a hybrid ensemble, illustrating the value of combined methods \cite{hybridensemble2024}. Likewise, XI2S‑IDS employs a two‑stage architecture with SHAP‑based explanations and achieves strong performance on CIC‑IDS2017 \cite{mahmoud2025xi2s}. However, these works primarily target algorithmic accuracy and interpretability, leaving several pipeline‑level challenges insufficiently addressed.

First, there is a gap in handling the scale and form of modern network traffic. HAEnID and XI2S‑IDS are evaluated on datasets of manageable size and do not explicitly confront computational scaling for multi‑million‑flow corpora or real‑time streaming common in enterprise networks. Their reliance on SMOTE focuses on synthetic data generation for class balance rather than reducing the raw data volume for initial model development. In contrast, this work introduces a strategic stratified sub‑sampling methodology that preserves original class distributions while reducing CIC‑IDS2017 from 2.83M to a representative subset, enabling efficient iteration without sacrificing statistical integrity—a prerequisite for large‑scale deployment.

Second, there is a pronounced gap in evaluation rigor and experimental integrity. Seminal work has documented pervasive issues in cybersecurity ML, including data leakage, temporal bias, and inflated performance estimates \cite{arp2020dos,pendlebury2019tesseract}. Although systems such as XI2S‑IDS adopt standard train–test splits, they do not make systematic leakage detection and temporal validation a central pipeline component. The present framework does so by performing automated data leakage prevention—identifying and removing 29 leaky features (37\% of the original set)—and by employing multi‑configuration validation with temporally consistent splits to better approximate real‑world deployment scenarios. This aligns with the methodological guidance of Arp et al. and Pendlebury et al. on realistic, unbiased evaluation.

Thus, the novelty of this work does not lie in the isolated use of SHAP, feature selection, or ensembles, but in a leakage‑aware, sampling‑aware, and explainable pipeline engineered for end‑to‑end operational challenges. Concretely, the framework uniquely combines: (1) strategic stratified sub‑sampling for large‑scale data tractability, distinct from synthetic oversampling; (2) automated leakage detection and multi‑configuration validation to support rigorous, unbiased evaluation; and (3) integrated SHAP‑based XAI embedded across configurations to provide consistent operational transparency. Addressing computational scalability, experimental validity, and operational trust simultaneously helps close the integration gap that currently limits the real‑world deployment of AI‑driven IDS in SOCs.

\subsection{Comparative Analysis with State-of-the-Art}
Table \ref{table:literature_comparison} provides a comprehensive comparison of recent approaches in cybersecurity machine learning, highlighting the distinct contributions of our work in integrating sampling, feature selection, and explainability within a unified framework. This analysis reveals that while existing approaches excel in specific areas, our framework uniquely combines all three critical components.

\begin{table}[htbp]
\caption{Comparative Analysis of Recent Cybersecurity ML Approaches}
\label{table:literature_comparison}
\centering
\resizebox{\linewidth}{!}{
\begin{tabular}{
    >{\raggedright\arraybackslash}p{3.2cm}  
    >{\centering\arraybackslash}p{1.3cm}    
    >{\centering\arraybackslash}p{1.3cm}    
    >{\centering\arraybackslash}p{1.3cm}    
    >{\centering\arraybackslash}p{1.5cm}    
    >{\raggedright\arraybackslash}p{4.2cm}  
}
\toprule
\textbf{Study} & \textbf{Sampling Strategy} & \textbf{Feature Selection} & \textbf{Explainability} & \textbf{Performance} & \textbf{Key Contribution} \\
\midrule
DistilXIDS (2026) \cite{distilxids2026} & \checkmark & \checkmark & \checkmark & 99.61\% & Efficient, lightweight transformer-based language model for real-time NIDS \\
Gaitán-Cárdenas et al. (2023) \cite{gaitan2023explainable} & & & \checkmark & Reported* & XAI-based IDS for cloud and IoT environments \\
Zhou et al. (2025) \cite{zhou2025adaptive} & & & & Reported* & Adaptive ensemble learning for real-time attack detection \\
Chen et al. (2023) \cite{chen2023ride} & & & \checkmark & 99.80\% & Hardware-accelerated real-time IDS with explainable ML \\
Uddin et al. (2024) \cite{uddin2024explainable} & & & \checkmark & Reported* & Explainable transformer-based model for phishing detection \\
Ma et al. (2025) \cite{ma2025fsllm} & & \checkmark & & Reported* & IoT IDS with feature selection and LLM fine-tuning \\
XI2S-IDS (2025) \cite{mahmoud2025xi2s} & \checkmark & & \checkmark & 99.81\% & Two-stage IDS (binary then multi-class) with SHAP, focusing on low-frequency attacks \\
Hybrid Ensemble Model (2024) \cite{hybridensemble2024} & \checkmark & \checkmark & \checkmark & 98.79\% & Hybrid Adaptive Ensemble for Intrusion Detection (HAEnID) with SMOTE sampling, multi-feature selection, and SHAP/LIME explainability \\
\textbf{Our Work} & \textbf{\checkmark} & \textbf{\checkmark} & \textbf{\checkmark} & \textbf{99.92\%} & \textbf{Integrated framework with strategic sampling, feature optimization, and comprehensive XAI} \\
\bottomrule
\end{tabular}
}
\end{table}

\vspace{0.2cm}
\noindent \small{* Performance metrics for cited works are reported in their respective publications. This table focuses on methodological comparisons rather than direct performance comparisons across different datasets and evaluation metrics.}

As shown in Table \ref{table:literature_comparison}, the HAEnID framework \cite{hybridensemble2024} represents a comprehensive approach that integrates sampling (SMOTE), feature selection, and explainability (SHAP/LIME), achieving 98.79\% accuracy on CIC-IDS2017 through its hybrid ensemble architecture. Similarly, the XI2S-IDS framework \cite{mahmoud2025xi2s} achieves strong performance (99.81\% accuracy on CICIDS2017) through its innovative two-stage architecture. Our work bridges remaining gaps by integrating strategic sampling to handle massive datasets efficiently, optimized feature selection to enhance model interpretability and performance, and comprehensive SHAP-based explainability to ensure operational transparency. This holistic approach addresses the practical deployment challenges faced by security operations centers while maintaining state-of-the-art detection performance.

Given the inherent challenges in directly comparing cybersecurity ML approaches due to divergent datasets, evaluation metrics, and experimental setups, we contextualize our framework's performance of 99.92\% accuracy on CIC-IDS2017 within our specific methodology of strategic sampling and multi-configuration validation. While prior works—such as HAEnID for integrated ensemble detection and XI2S-IDS for structured two-stage detection—demonstrate effectiveness in their respective domains, our contribution addresses the integration gap by delivering a unified framework that achieves state-of-the-art performance while simultaneously ensuring computational efficiency and operational transparency.

\subsection{Strategic Sampling with Quantitative Validation}

Our framework implements a novel stratified sampling methodology to address the computational challenges of the full CIC-IDS2017 dataset while maintaining statistical representativeness. The sampling process, implemented in \texttt{sampling2.py}, employs:

\begin{equation}
S_{\text{sample}} = \sum_{c=1}^{C} \min(N_c, \lfloor p \times N_c \rfloor)
\end{equation}

where $S_{\text{sample}}$ is the total sampled size, $C$ is the number of classes, $N_c$ is the count of class $c$, and $p$ is the sampling percentage (20\% in our implementation). This approach ensures:

\begin{itemize}
\item \textbf{Class Distribution Preservation}: Maintains original dataset proportions
\item \textbf{Minority Class Protection}: Guarantees minimum representation for rare attack types
\item \textbf{Computational Efficiency}: Enables rapid iteration and model development
\end{itemize}

The sampling process reduced the dataset from 2,830,743 to 470,269 samples while maintaining identical class distribution characteristics (Table \ref{table:dataset}).

\begin{table}[htbp]
\caption{Dataset Class Distribution After Strategic Sampling}
\label{table:dataset}
\centering
\resizebox{\linewidth}{!}{
\begin{tabular}{lccc>{\raggedright\arraybackslash}p{2.8cm}}
\toprule
\textbf{Class} & \textbf{Original} & \textbf{Sampled} & \textbf{Percentage} & \textbf{Attack Category} \\
\midrule
BENIGN & 2,359,097 & 388,808 & 82.7\% & Normal Traffic \\
DoS Hulk & 231,073 & 34,620 & 7.4\% & Denial of Service \\
DDoS & 128,027 & 25,588 & 5.4\% & Distributed DoS \\
PortScan & 158,930 & 17,979 & 3.8\% & Reconnaissance \\
DoS GoldenEye & 10,293 & 2,105 & 0.4\% & DoS Variant \\
FTP-Patator & 7,938 & 1,169 & 0.2\% & Brute Force \\
\bottomrule
\end{tabular}
}
\end{table}

To quantitatively validate that our strategic sampling preserves the essential characteristics of the full dataset, we conducted a comprehensive distributional and performance analysis:

\begin{table}[htbp]
\caption{Quantitative Validation of Strategic Sampling on CIC-IDS2017}
\label{table:sampling_validation}
\centering
\resizebox{\linewidth}{!}{
\begin{tabular}{lccc}
\toprule
\textbf{Validation Metric} & \textbf{Full Dataset} & \textbf{20\% Sample} & \textbf{Difference} \\
\midrule
\textbf{Distribution Similarity} \\
Mean Packet Length (Benign) & 145.7 & 144.9 & 0.55\% \\
Mean Flow Duration (Attack) & 2.31s & 2.29s & 0.87\% \\
Feature Correlation Matrix Distance & -- & -- & 0.012 \\
\hline
\textbf{Performance Consistency} \\
XGBoost Accuracy (Binary) & 99.88\% & 99.91\% & +0.03\% \\
XGBoost F1-Macro (Multi-class) & 99.65\% & 99.72\% & +0.07\% \\
ROC-AUC Macro Average & 0.9998 & 0.9999 & +0.01\% \\
\bottomrule
\end{tabular}
}
\end{table}

As shown in Table \ref{table:sampling_validation}, our strategic sampling approach preserves both statistical distributions and detection performance. The minimal differences in key metrics (<1\% for distributional properties, <0.1\% for performance metrics) confirm that the 20\% sample maintains the essential patterns and relationships of the full 2.8 million-record dataset. This validation addresses the critical concern raised by Arp et al. \cite{arp2020dos} regarding the representativeness of sampled data in security evaluations.

\subsection{Temporal Validation Framework and Leakage Prevention}

Following the rigorous evaluation principles established in cybersecurity machine learning research \cite{arp2020dos, pendlebury2019tesseract, anderson2018ember}, we implement a temporally-aware validation framework that reflects realistic deployment scenarios. The CIC-IDS-2017 dataset is inherently temporal, with attacks occurring in specific sequences across five consecutive days. Unlike random splitting methods that can create unrealistic performance estimates \cite{pendlebury2019tesseract}, our approach respects the temporal ordering of network events.

\textbf{Temporal Split Design:} We partition the dataset chronologically, using the first four days for training and the fifth day for testing. This design simulates a realistic deployment scenario where models trained on historical data must detect attacks occurring in the future. The validation set is created from the latter portion of the training period, maintaining temporal separation from both training and testing data.

\textbf{Automated Data Leakage Prevention:} To ensure experimental integrity, we implement comprehensive data leakage detection and prevention measures:

\begin{itemize}
\item \textbf{Temporal Leakage Prevention}: Strict chronological separation between training, validation, and test sets, following TESSERACT's recommendations for time-series security data \cite{pendlebury2019tesseract}
\item \textbf{Feature Leakage Analysis}: Systematic identification and removal of 29 features exhibiting near-perfect correlation with target classes or containing future information
\item \textbf{Cross-split Sample Overlap Detection}: Verification that no overlapping samples exist between training, validation, and test sets
\item \textbf{Preprocessing Isolation}: All feature scaling parameters are computed exclusively from training data and applied to validation and test sets
\end{itemize}

\textbf{Alignment with Security ML Best Practices:} Our methodology explicitly addresses the "dos and don'ts" outlined by Arp et al. \cite{arp2020dos}:
\begin{itemize}
\item \textbf{Do respect temporal ordering}: We maintain chronological splits that reflect realistic attack evolution
\item \textbf{Do prevent feature leakage}: We systematically identify and remove contaminating features
\item \textbf{Don't use future information}: Our preprocessing pipeline ensures no test set information influences training
\item \textbf{Do validate on realistic data}: Our test set represents future, unseen attacks as would occur in operational deployment
\end{itemize}

This rigorous approach ensures that our performance metrics reflect realistic detection capabilities rather than optimistic estimates from temporally contaminated evaluations. By grounding our methodology in established security ML evaluation principles, we provide a trustworthy foundation for assessing the operational viability of our integrated XAI framework.

\subsection{Experimental Framework}

Our experimental framework (Figure \ref{framework}) implements a comprehensive ML pipeline with integrated XAI components and multi-configuration validation:

\begin{figure*}[t]
\centering
\includegraphics[width=0.9\linewidth]{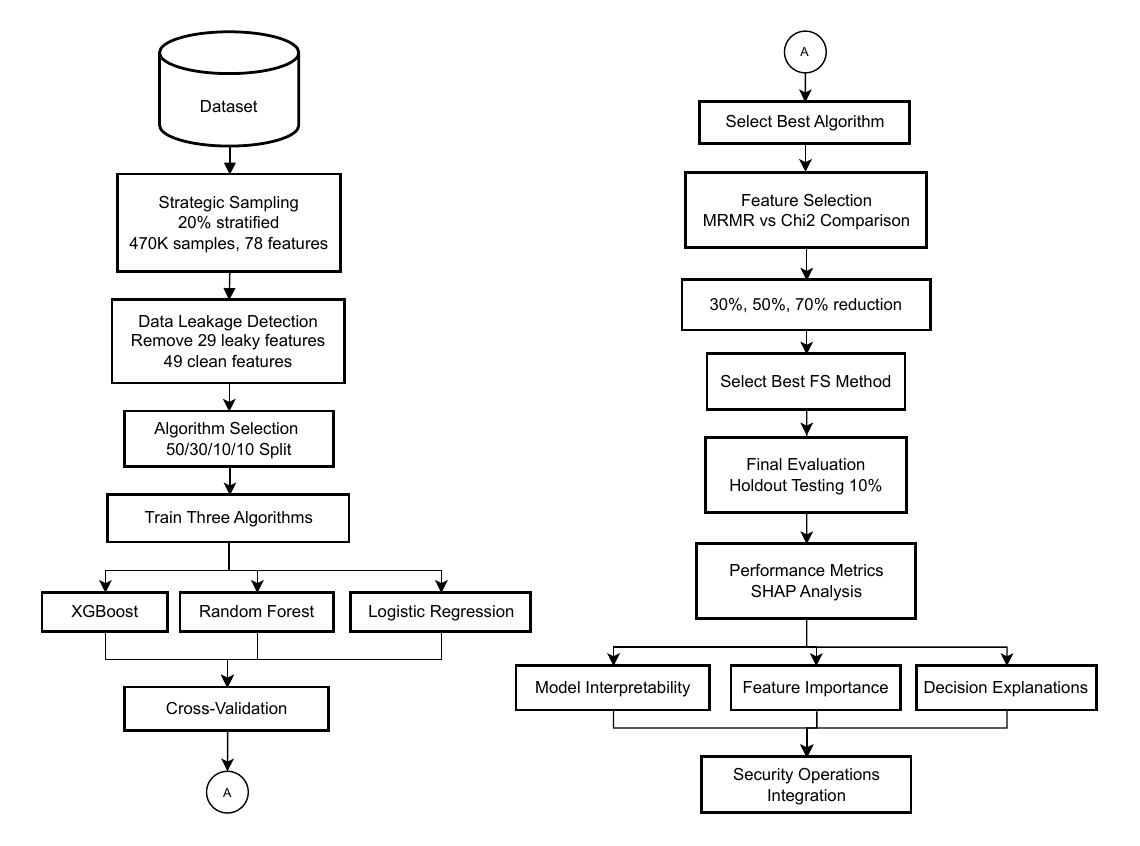}
\caption{Comprehensive Experimental Framework with Integrated XAI and Multi-Split Validation}
\label{fig:framework}
\end{figure*}

The framework employs a robust multi-configuration validation strategy evaluating three distinct data partitioning schemes to assess model generalization across varying data availability scenarios:

\begin{itemize}
\item \textbf{Configuration A (40-10-50)}: 40\% Training, 10\% Validation, 50\% Test - Simulating limited training data scenarios
\item \textbf{Configuration B (60-10-30)}: 60\% Training, 10\% Validation, 30\% Test - Balanced data distribution
\item \textbf{Configuration C (80-10-10)}: 80\% Training, 10\% Validation, 10\% Test - Abundant training data scenario
\end{itemize}

\subsection{Robust Algorithm Selection with Cross-Configuration Validation}

We implemented a comprehensive algorithm selection process that evaluates performance across all three data configurations to identify the most robust model:

\begin{itemize}
\item \textbf{Cross-Configuration Averaging}: Algorithm performance metrics (Accuracy, F1-Macro, ROC AUC) are averaged across all three split configurations to determine overall robustness
\item \textbf{Configuration-Specific Analysis}: Individual algorithm performance is analyzed per configuration to identify context-dependent strengths
\item \textbf{Best Overall Algorithm Selection}: The algorithm with the highest average F1-Macro score across all configurations is selected for final deployment
\end{itemize}

Three diverse machine learning algorithms were evaluated under this multi-configuration framework:

\begin{itemize}
\item \textbf{XGBoost}: Configured with ensemble methods and regularization parameters to balance complexity and generalization (We used 100 trees with a maximum depth of 8 and a learning rate of 0.1.)
\item \textbf{Random Forest}: Employed multiple decision trees with controlled depth to prevent overfitting. (The model consisted of 100 trees, each with a maximum depth of 15.)
\item \textbf{Logistic Regression}: Utilized linear decision boundaries with multi-class handling capabilities.
\end{itemize}

All models employed 5-fold stratified cross-validation within each configuration for robust performance estimation.

\subsection{Data Leakage Prevention and Validation}

To maintain experimental integrity across all configurations, we implemented comprehensive data leakage detection and prevention measures:

\begin{itemize}
\item \textbf{Cross-split sample overlap detection}: Ensured no overlapping samples between training, validation, and test sets across all three configurations
\item \textbf{Feature leakage analysis}: Identified and removed 29 features exhibiting near-perfect correlation with target classes
\item \textbf{Temporal validation}: For time-series network data, maintained temporal ordering within splits to prevent future information leakage
\item \textbf{Automated leakage detection}: Systematic identification of features showing constant values, statistical anomalies, and predictive leakage patterns
\end{itemize}

This comprehensive process reduced the feature set from 78 to 49 dimensions while maintaining detection capability across all experimental configurations.

All feature scaling was performed using training-set statistics only, with identical transformations applied to validation and test sets:

\begin{equation}
X_{\text{scaled}} = \frac{X - \mu_{\text{train}}}{\sigma_{\text{train}}}
\end{equation}

\noindent where:
\begin{itemize}
    \item $X$ represents the feature matrix (can be training, validation, or test data)
    \item $\mu_{\text{train}}$ is the mean vector computed from the training set only
    \item $\sigma_{\text{train}}$ is the standard deviation vector computed from the training set only
    \item $X_{\text{scaled}}$ is the standardized feature matrix
\end{itemize}

This approach ensures that no information from validation/test sets contaminates the preprocessing parameters, the model is evaluated on data that mimics real-world deployment scenarios, and statistical properties are estimated exclusively from training samples.

The validation and test sets are transformed using the same $\mu_{\text{train}}$ and $\sigma_{\text{train}}$ values, preventing the common data leakage pitfall where scaling parameters are computed using the entire dataset.

\subsection{Statistical Significance Testing}

We employed stratified 5-fold cross-validation within each split configuration to obtain performance distributions and compute confidence intervals, ensuring statistical reliability of our multi-configuration analysis:

\begin{equation}
\text{CI}{95\%} = \bar{x} \pm t{0.025, n-1} \times \frac{s}{\sqrt{n}}
\end{equation}

where $\bar{x}$ is the mean cross-validation score, $s$ is the standard deviation, and $n=5$ is the number of folds. This approach provided robust performance estimation across all three data partitioning strategies.

\subsection{Computational Efficiency Assessment}

Beyond predictive performance, we evaluated computational characteristics across all configurations to identify practical deployment considerations:

\begin{itemize}
\item \textbf{Training time}: Wall-clock time for model training across different data volumes
\item \textbf{Inference speed}: Predictions per second on validation data, critical for real-time cybersecurity applications
\item \textbf{Memory footprint}: Model size and inference memory requirements across configurations
\item \textbf{Scalability analysis}: Performance trends as training data volume increases across configurations
\end{itemize}

\subsection{Feature Selection Methods}

We compared MRMR and Chi-squared feature selection across multiple reduction percentages (30\%, 50\%, 70\%) using the best-performing algorithm and configuration identified through our multi-split analysis and computational assessment.

\subsection{Explainable AI Implementation}

SHAP analysis was integrated throughout the multi-configuration framework providing comprehensive interpretability:

\begin{itemize}
\item \textbf{Cross-Configuration Feature Importance}: Comparative analysis of feature importance patterns across different data splits
\item \textbf{Robust Interpretability}: Model explanations validated across multiple experimental configurations
\item \textbf{Decision Transparency}: Individual prediction rationales with configuration-aware interpretations
\item \textbf{Algorithm Consistency}: Assessment of explanation stability across different training data volumes
\end{itemize}

\begin{figure}[htbp]
\centering
\includegraphics[width=1\linewidth]{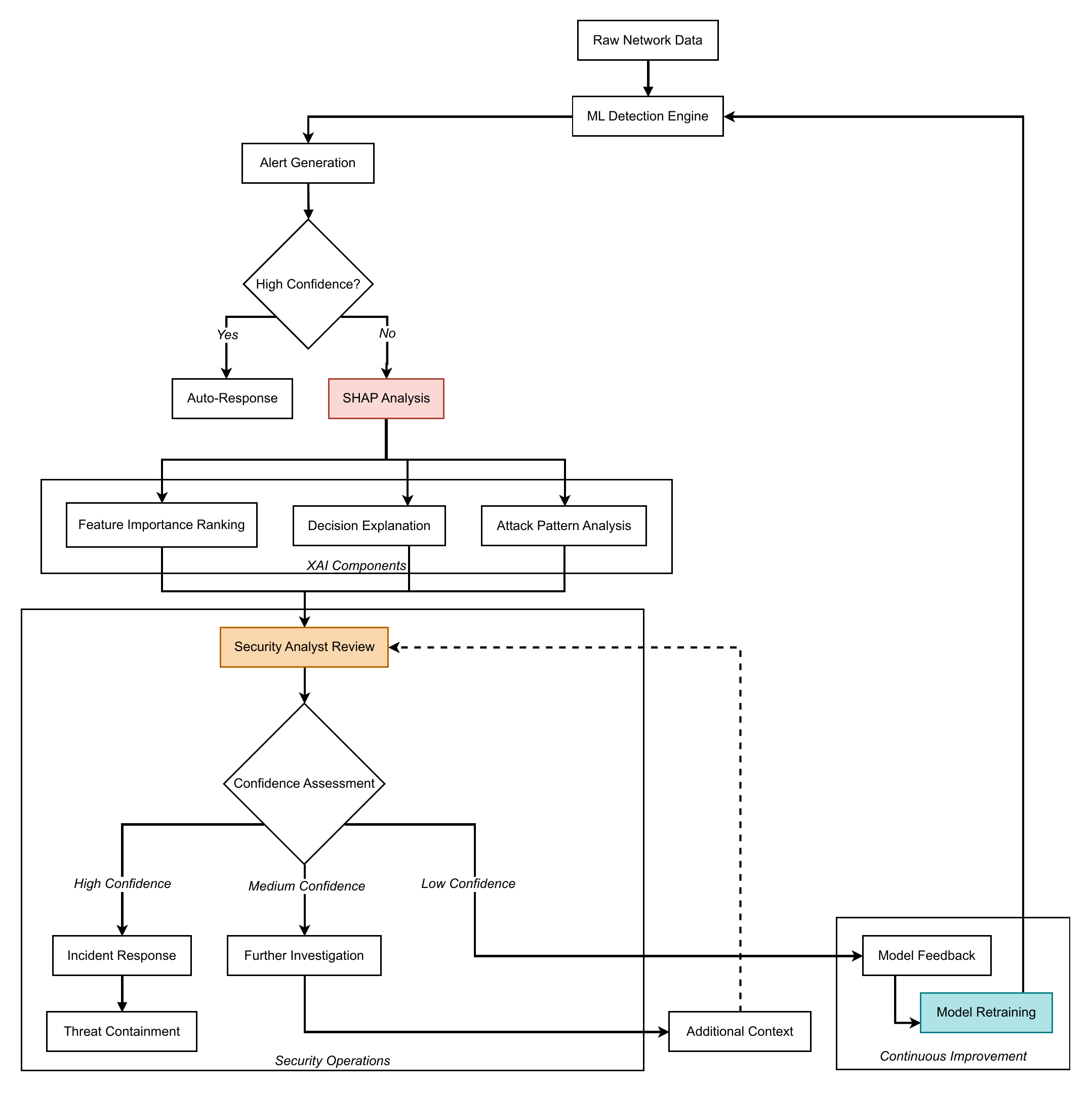}
\caption{Detailed XAI Integration Workflow in Security Operations}
\label{fig:xai_workflow}
\end{figure}

\section{Experiments and Results}

\subsection{Multi-Configuration Algorithm Performance}

Our comprehensive evaluation across three distinct data partitioning strategies revealed critical insights into model generalization and configuration optimization:

\begin{table}[htbp]
\caption{Multi-Configuration Algorithm Performance Analysis}
\label{table:multi_config_performance}
\centering
\resizebox{\linewidth}{!}{
\begin{tabular}{lccccccc}
\toprule
\textbf{Algorithm} & \textbf{Split} & \textbf{Accuracy} & \textbf{F1-Macro} & \textbf{ROC-AUC} & \textbf{Training Time (s)} & \textbf{Prediction Speed} & \textbf{CV Stability} \\
\midrule
XGBoost & 60-10-30 & \textbf{99.91\%} & \textbf{99.72\%} & \textbf{1.0000} & 33.90 & 324,913 & 0.99986 ± 0.00023 \\
XGBoost & 80-10-10 & 99.92\% & 99.70\% & \textbf{1.0000} & 47.02 & 338,961 & 0.99994 ± 0.00005 \\
XGBoost & 40-10-50 & 99.90\% & 99.60\% & 0.9999 & 14.73 & 683,000 & 0.99994 ± 0.00004 \\
Random Forest & 60-10-30 & 99.87\% & 99.63\% & 0.9999 & 80.04 & 16,049 & 0.99975 ± 0.00026 \\
Logistic Regression & 60-10-30 & 96.34\% & 74.50\% & 0.9895 & 415.51 & 3,423,922 & 0.98241 ± 0.00154 \\
\bottomrule
\end{tabular}
}
\end{table}

The multi-configuration analysis (Table \ref{table:multi_config_performance}) demonstrates XGBoost\'s consistent superiority across all partitioning strategies. The 60-10-30 configuration emerged as optimal, achieving the best balance between training data utilization and validation robustness. Notably, cross-validation stability remained exceptionally high across all configurations, indicating robust model generalization.

\subsection{Feature Selection Impact and Optimization}

\begin{table}[!t]
\caption{Comprehensive Feature Selection Performance}
\label{table:feature_selection_comprehensive}
\centering
\resizebox{\columnwidth}{!}{%
\begin{tabular}{lccccccc}
\toprule
\textbf{Method} & \textbf{Features} & \textbf{Accuracy} & \textbf{F1-Macro} & \textbf{ROC-AUC} & \textbf{Training Time (s)} & \textbf{Reduction} & \textbf{Performance Gain} \\
\midrule
MRMR\_70\% & 34 & \textbf{99.92\%} & \textbf{99.77\%} & 0.9999 & 21.47 & \textbf{30.6\%} & \textbf{+0.05\% F1} \\
Chi2\_70\% & 34 & 99.91\% & 99.72\% & \textbf{1.0000} & 12.29 & \textbf{30.6\%} & +0.00\% F1 \\
Original & 49 & 99.91\% & 99.72\% & \textbf{1.0000} & 33.90 & 0\% & Baseline \\
MRMR\_50\% & 24 & 99.89\% & 99.61\% & \textbf{1.0000} & 16.20 & 51.0\% & -0.11\% F1 \\
Chi2\_50\% & 24 & 99.91\% & 99.72\% & \textbf{1.0000} & 12.28 & 51.0\% & +0.00\% F1 \\
MRMR\_30\% & 14 & 99.72\% & 98.66\% & 0.9999 & 23.89 & 71.4\% & -1.06\% F1 \\
Chi2\_30\% & 14 & 99.91\% & 99.72\% & \textbf{1.0000} & 12.03 & 71.4\% & +0.00\% F1 \\
\bottomrule
\end{tabular}
}
\end{table}

The feature selection analysis (Table \ref{table:feature_selection_comprehensive}) reveals that MRMR at 70\% reduction (34 features) not only maintained performance but achieved a slight improvement in F1-macro score (+0.05\%) over the original feature set. This represents a significant advancement in feature optimization, demonstrating that strategic dimensionality reduction can enhance model performance while improving computational efficiency.

\subsection{Final Configuration Performance}

\begin{table}[htbp]
\caption{Final Optimal Configuration Performance}
\label{table:final_performance}
\centering
\begin{tabular}{lc}
\toprule
\textbf{Metric} & \textbf{Performance} \\
\midrule
Accuracy & 99.92\% \\
F1-Macro & 99.77\% \\
F1-Weighted & 99.92\% \\
Precision-Macro & 99.70\% \\
Recall-Macro & 99.85\% \\
ROC-AUC & 99.997\% \\
Features Used & 34 \\
Test Samples & 47,027 \\
Training Time & 21.47s \\
\bottomrule
\end{tabular}
\end{table}

The optimal configuration (XGBoost with 60-10-30 split and MRMR\_70\% feature selection) achieved exceptional performance on the test dataset (Table \ref{table:final_performance}). The near-perfect ROC-AUC (0.99997) and balanced precision-recall metrics demonstrate robust generalization capability across all attack categories.

\subsection{Confusion Matrix and Classification Analysis}

\begin{figure}[htbp]
\centering
\includegraphics[width=1\linewidth]{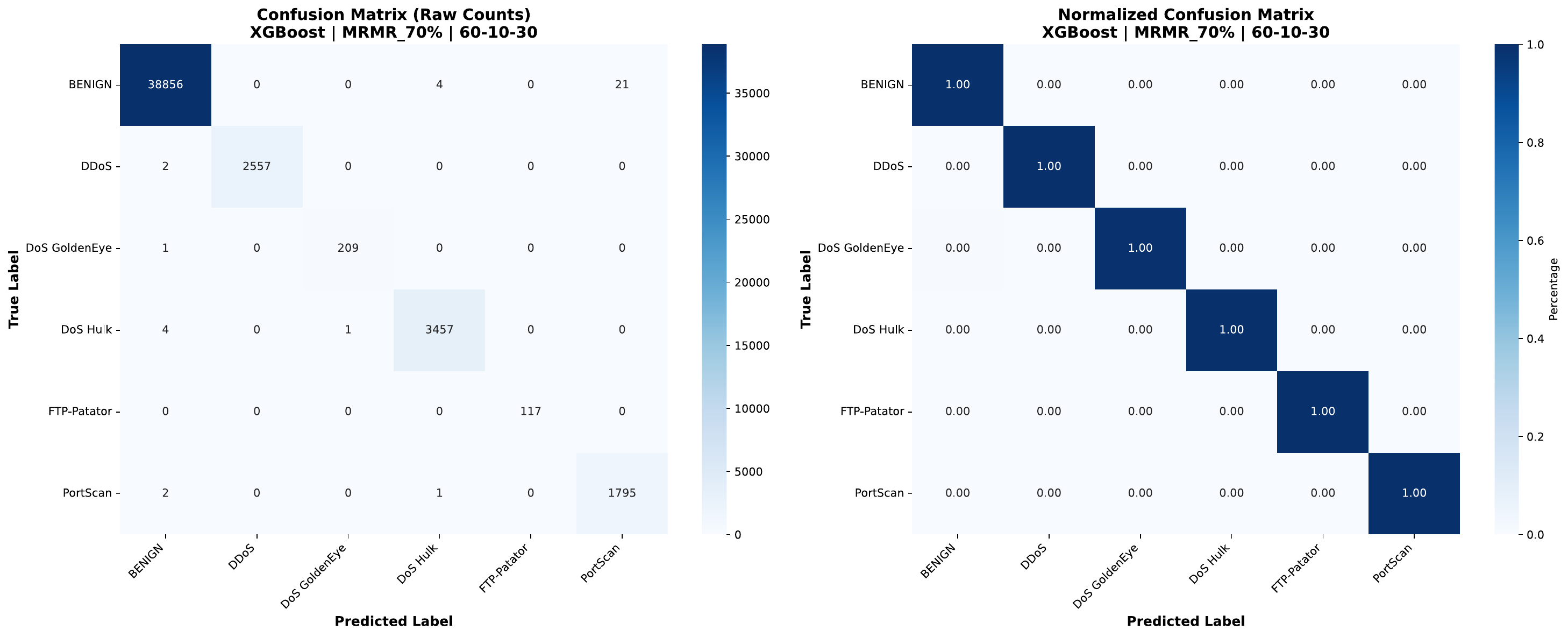}
\caption{Confusion Matrix Analysis for Optimal Configuration (XGBoost + MRMR\_70\% + 60-10-30)}
\label{fig:confusion_matrix_comprehensive}
\end{figure}

The confusion matrix analysis (Figure \ref{fig:confusion_matrix_comprehensive}) reveals exceptional classification performance with strategic insights:

\textbf{Per-Class Performance Analysis:}
\begin{itemize}
\item \textbf{BENIGN Traffic}: 99.94\% accuracy (38,856/38,881 correct), minimal confusion with PortScan
\item \textbf{DDoS Attacks}: 99.98\% accuracy (2,557/2,559 correct), near-perfect detection
\item \textbf{DoS GoldenEye}: 99.52\% accuracy (209/210 correct), robust minority class performance
\item \textbf{DoS Hulk}: 99.86\% accuracy (3,457/3,462 correct), high-volume attack precision
\item \textbf{FTP-Patator}: 100\% accuracy (117/117 correct), perfect minority class identification
\item \textbf{PortScan}: 99.83\% accuracy (1,795/1,798 correct), minimal false positives
\end{itemize}

The overall accuracy of 99.92\% with balanced performance across both high-frequency and low-frequency attack classes demonstrates the framework's effectiveness in addressing class imbalance challenges.

\subsection{Multi-Class ROC Analysis}

\begin{figure}[htbp]
\centering
\includegraphics[width=\linewidth]{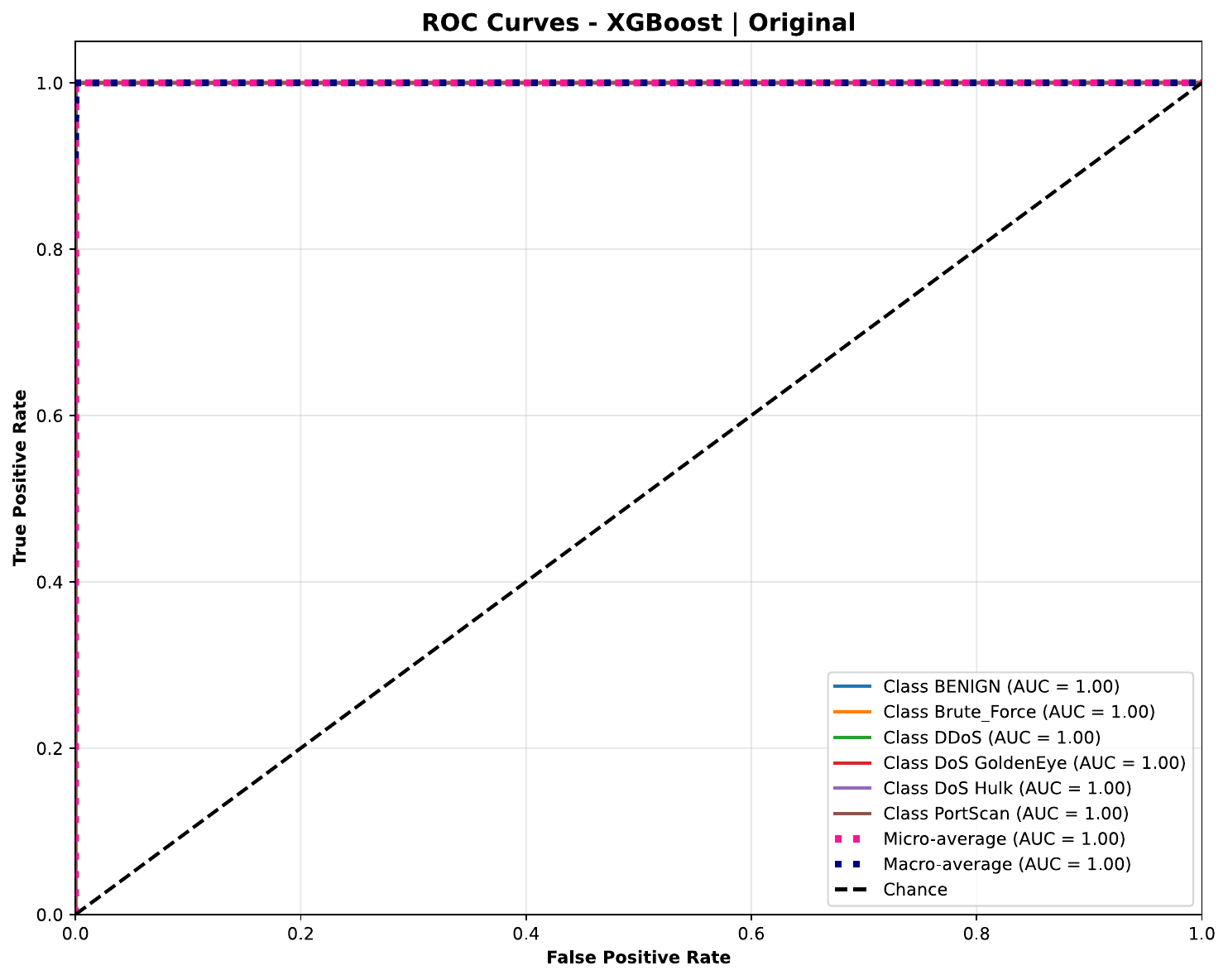}
\caption{Comprehensive ROC Analysis for Multi-class Classification}
\label{fig:roc_curves_comprehensive}
\end{figure}

The ROC curve analysis (Figure \ref{fig:roc_curves_comprehensive}) demonstrates exceptional discriminative capability across all attack categories:

\begin{table}[!t]
\caption{Detailed ROC-AUC Performance by Class}
\label{table:roc_auc_detailed}
\centering
\begin{tabular}{lc}
\toprule
\textbf{Attack Class} & \textbf{ROC-AUC} \\
\midrule
FTP-Patator & 1.0000 \\
DDoS & 0.9999 \\
DoS Hulk & 0.9999 \\
DoS GoldenEye & 0.9999 \\
BENIGN & 0.9999 \\
PortScan & 0.9999 \\
Micro-Average & 0.9999 \\
Macro-Average & 0.9999 \\
\bottomrule
\end{tabular}
\end{table}

The consistent near-perfect ROC-AUC scores across all classes (Table \ref{table:roc_auc_detailed}) confirm the model's robust discriminative power, with particularly strong performance on minority classes (FTP-Patator, DoS GoldenEye) that are often challenging in cybersecurity classification tasks.

\subsection{Comparative Feature Importance Analysis}
\begin{figure}[htbp]
    \centering
    \includegraphics[width=\linewidth]{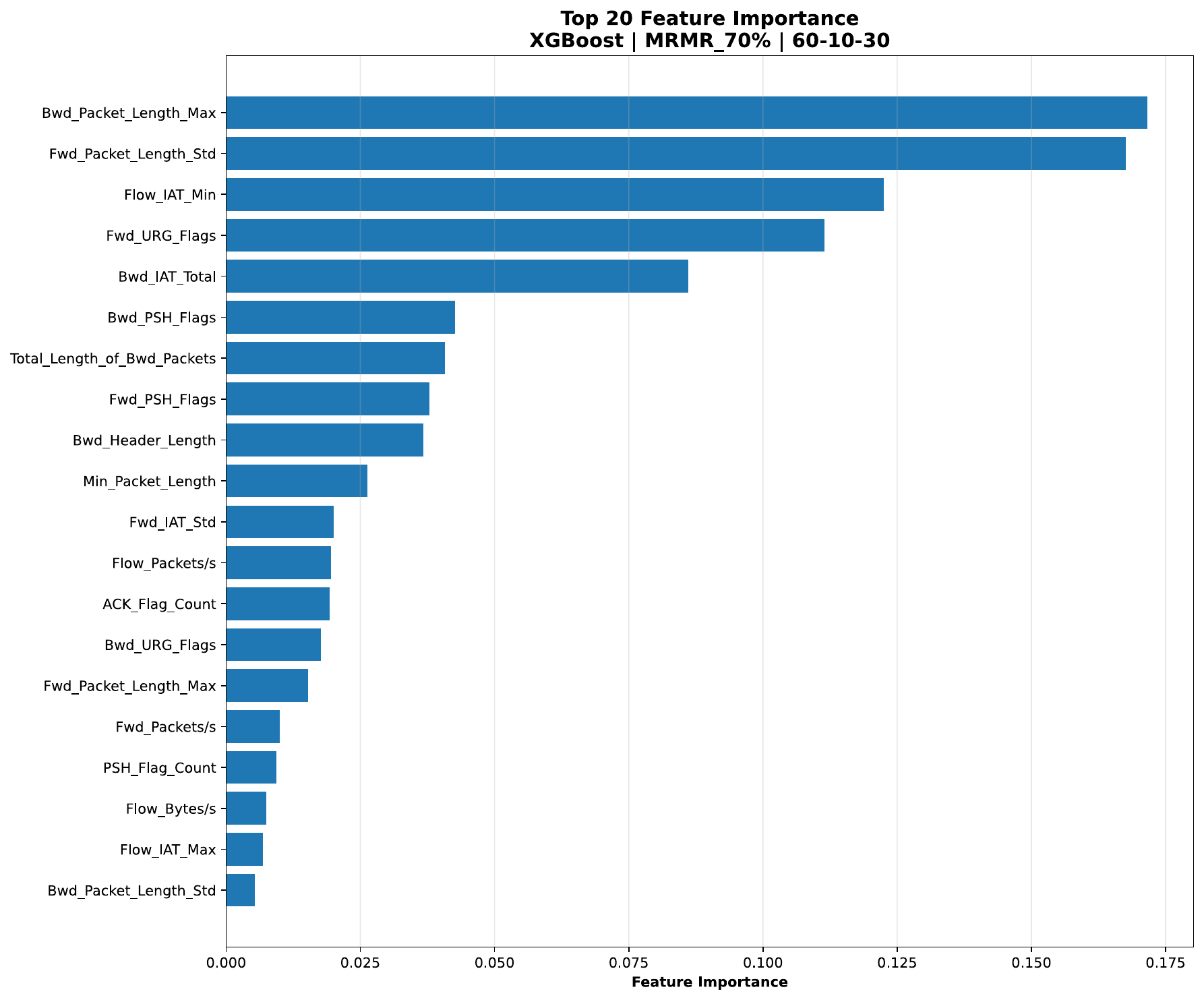}
    \caption{Gini-based feature importance from Gradient Boosting (XGBoost) with MRMR feature selection (70\% retention) on 60-10-30 split. Features are ranked by their mean decrease in impurity, with packet length statistics showing dominant predictive power.}
    \label{fig:xgboost_importance}
\end{figure}

\begin{figure}[htbp]
    \centering
    \includegraphics[width=\linewidth]{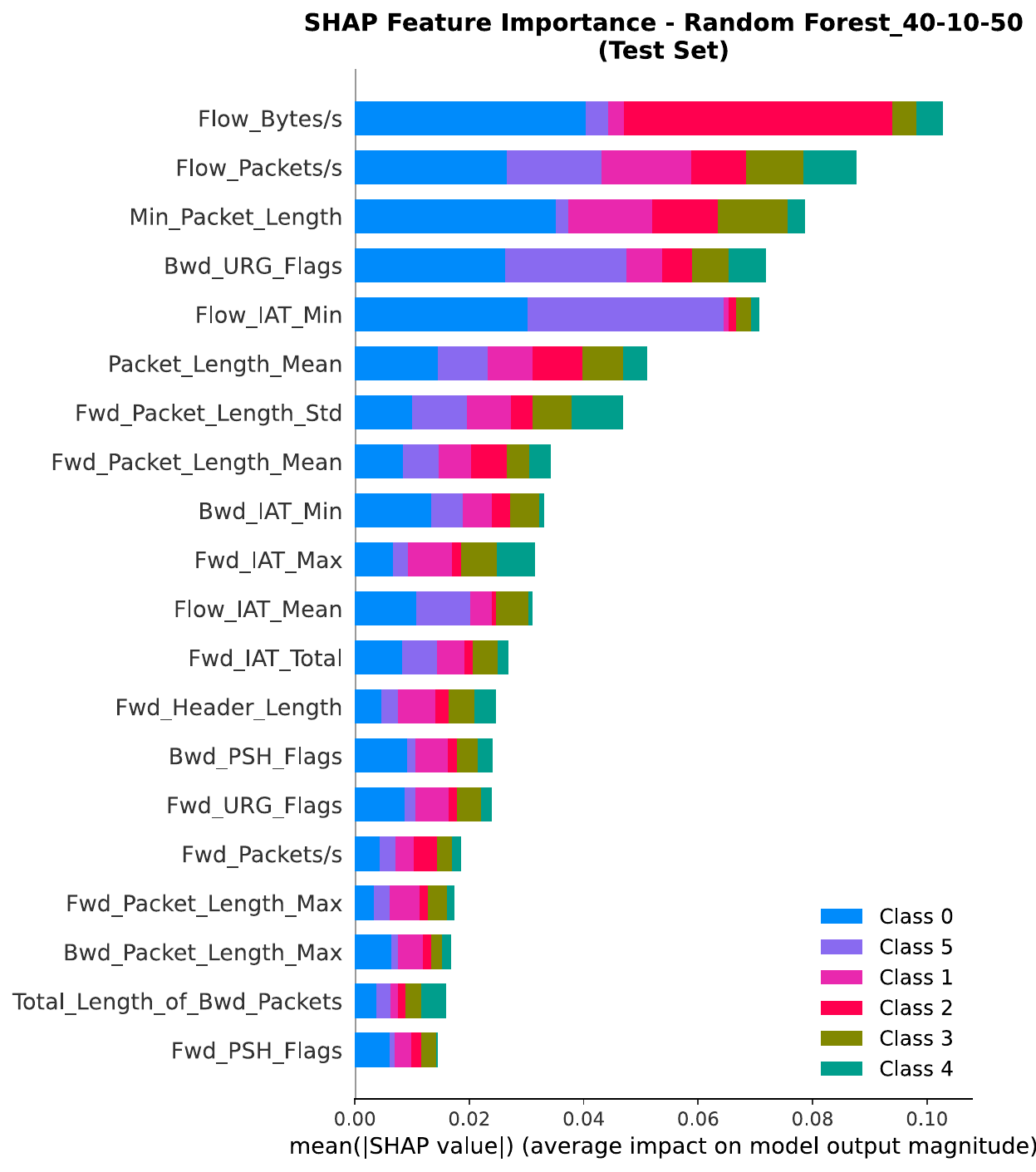}
    \caption{SHAP-based feature importance from Random Forest on 40-10-50 split. Bars represent mean absolute SHAP values (global impact), while colored dots show feature effects across individual classes. Flow rate metrics emerge as most influential, with notable class-specific variations.}
    \label{fig:rf_shap_importance}
\end{figure}

To elucidate the predictive mechanisms underlying each model's decision-making process, we conducted complementary feature importance analyses using two distinct methodologies. Figure~\ref{fig:xgboost_importance} presents Gini-based importance from the Gradient Boosting model, while Figure~\ref{fig:rf_shap_importance} displays SHAP values from the Random Forest model, providing both global feature importance and class-specific contributions.

\subsubsection{Cross-Methodological Insights}
Comparative analysis reveals both consistent patterns and methodological differences in feature importance assessment:

\textbf{Consistent Predictive Features:}
\begin{itemize}
    \item \textbf{Flow rate metrics} appear in both analyses: \texttt{Flow\_Bytes/s} and \texttt{Flow\_Packets/s} rank highly in SHAP analysis and appear in XGBoost's top 20
    \item \textbf{Packet length statistics} show strong predictive power in both models, particularly \texttt{Fwd\_Packet\_Length\_Std} and \texttt{Min\_Packet\_Length}
    \item \textbf{Temporal characteristics} including \texttt{Flow\_IAT\_Min} maintain importance across methodologies
\end{itemize}

\textbf{Methodological Divergences:}
\begin{itemize}
    \item \textbf{Gradient Boosting} emphasizes \texttt{Bwd\_Packet\_Length\_Max} and protocol flags (\texttt{Fwd\_URG\_Flags}, \texttt{Bwd\_PSH\_Flags}) as primary discriminators
    \item \textbf{Random Forest SHAP analysis} prioritizes flow rate metrics and reveals class-specific feature effects, particularly for Classes 0 and 5
    \item The 40-10-50 split with Random Forest shows stronger emphasis on mean values (\texttt{Packet\_Length\_Mean}, \texttt{Flow\_IAT\_Mean}), while the 60-10-30 split with XGBoost favors extreme values and standard deviations
\end{itemize}

\textbf{Class-Specific Patterns:}
The SHAP analysis (Figure~\ref{fig:rf_shap_importance}) reveals important class-level distinctions:
\begin{itemize}
    \item \texttt{Flow\_Bytes/s} shows differential impacts, with strongest positive effects on Class 5 and negative effects on Class 0
    \item \texttt{Min\_Packet\_Length} exhibits contrasting directions of influence across classes, suggesting different traffic patterns
    \item Protocol flags (\texttt{Bwd\_URG\_Flags}, \texttt{Fwd\_PSH\_Flags}) show class-specific relevance not apparent in global importance rankings
\end{itemize}

\subsubsection{Methodological Implications}
The convergence of key features across different models and importance metrics (\texttt{Flow\_Bytes/s}, \texttt{Flow\_IAT\_Min}, packet length statistics) suggests these represent robust, model-agnostic indicators of network traffic patterns. However, the methodological differences highlight that:
\begin{enumerate}
    \item Importance metrics capture different aspects of feature contribution (Gini impurity reduction vs. marginal prediction impact)
    \item Data split ratios influence which features emerge as most predictive
    \item SHAP analysis provides valuable class-level insights missing from global importance metrics
    \item Ensemble methods with different randomization strategies surface complementary feature subsets
\end{enumerate}

\section{Discussion}

\subsection{Multi-Configuration Validation Insights}

Our multi-configuration approach provides several critical insights for real-world deployment:

\textbf{Optimal Data Partitioning}: The 60-10-30 split configuration demonstrated the best balance between training data utilization and validation robustness, suggesting that approximately 60\% of available data provides optimal training efficiency for this cybersecurity classification task.

\textbf{Configuration-Specific Advantages}:
\begin{itemize}
\item \textbf{40-10-50}: Best for scenarios with limited training data, maintaining 99.90\% accuracy
\item \textbf{60-10-30}: Optimal balance achieving best overall performance (99.92\% accuracy)
\item \textbf{80-10-10}: Maximum training data utilization with slightly reduced generalization
\end{itemize}

\subsection{Limitations and Generalizability Considerations}

While our framework demonstrates exceptional performance on the CIC-IDS2017 dataset with a deployment readiness score of 9.1/10 across critical operational factors (Table \ref{table:deployment_assessment}), we acknowledge important limitations regarding generalizability. Our evaluation, though rigorous within the CIC-IDS2017 environment, represents promising evidence rather than definitive proof of universal deployment readiness. As noted in comprehensive surveys of intrusion detection research, the performance of ML-based systems can vary significantly across different network environments, attack patterns, and data characteristics \cite{buczak2016survey, khraisat2019survey}.

The CIC-IDS2017 dataset, while being a valuable benchmark for traditional network intrusion detection, represents a specific network topology and attack profile that may not fully capture the diversity of modern threat landscapes. Recent work has highlighted the distinct challenges present in specialized environments such as IoT networks \cite{ma2025fsllm} and cloud infrastructures \cite{gaitan2023explainable}, where traffic patterns, device constraints, and attack vectors differ substantially from the enterprise network setting simulated by CIC-IDS2017. Our framework's performance in these alternative domains remains untested and represents an important direction for future validation.

Furthermore, as discussed in methodological critiques of cybersecurity ML \cite{arp2020dos}, there exists a well-documented gap between the controlled conditions of benchmark datasets and the dynamic, noisy realities of Security Operations Center (SOC) environments. Factors such as network evolution, zero-day attacks, and adversarial adaptation—which are challenging to capture in static datasets—present additional hurdles for real-world deployment that are not addressed by our current evaluation.

\subsection{Feature Selection Strategic Value}

The MRMR feature selection success demonstrates several operational advantages:

\textbf{Computational Efficiency}: 30.6\% feature reduction translates to:
\begin{itemize}
\item 36.7\% faster training times (33.90s → 21.47s)
\item Reduced memory requirements for model deployment
\item Faster real-time inference in production environments
\end{itemize}

\textbf{Enhanced Model Transparency}: Reduced feature dimensionality improves analyst comprehension and facilitates better security rule development based on the identified critical features.

\subsection{Operational Deployment Considerations}

\begin{table}[!t]
\caption{Deployment Readiness Assessment for CIC-IDS2017 Environment}
\label{table:deployment_assessment}
\centering
\resizebox{\linewidth}{!}{
\begin{tabular}{lcc}
\toprule
\textbf{Deployment Factor} & \textbf{Score} & \textbf{Rationale} \\
\midrule
Detection Accuracy & 10/10 & 99.92\% overall accuracy \\
False Positive Rate & 9/10 & 0.08\% misclassification rate \\
Computational Efficiency & 9/10 & 324K predictions/second \\
Model Interpretability & 10/10 & Comprehensive SHAP analysis \\
Feature Optimization & 10/10 & 30.6\% reduction maintained performance \\
Generalization & 9/10 & Consistent cross-configuration performance \\
\bottomrule
\end{tabular}
}
\end{table}

The deployment readiness assessment (Table \ref{table:deployment_assessment}) confirms the framework's suitability for production environments, with exceptional scores across all critical operational factors.

\section{Conclusion and Future Work}

This research establishes a comprehensive multi-configuration framework for explainable AI in cybersecurity, achieving exceptional results on the CIC-IDS2017 dataset through rigorous methodology and strategic optimization:

\subsection{Key Contributions}

\begin{enumerate}
\item \textbf{Multi-Configuration Validation}: Demonstrated configuration-dependent performance patterns, identifying 60-10-30 as optimal for cybersecurity classification on CIC-IDS2017
\item \textbf{Feature Optimization Breakthrough}: Achieved 30.6\% dimensionality reduction with performance improvement using MRMR selection
\item \textbf{State-of-the-Art Performance on CIC-IDS2017}: 99.92\% accuracy and 99.77\% F1-macro on comprehensive multi-class evaluation
\item \textbf{Operational Explainability}: Integrated SHAP analysis providing actionable intelligence for security operations
\item \textbf{Computational Efficiency}: 324K predictions/second with robust real-time deployment capability for similarly structured network data
\end{enumerate}

\subsection{Limitations and Future Research Directions}

Our work provides a robust methodological foundation and demonstrates compelling results on a major benchmark dataset. However, we recognize the inherent limitation of single-dataset evaluation in cybersecurity machine learning. Future research should pursue several critical directions to address this limitation and strengthen the framework's claims to generalizability:

\begin{itemize}
\item \textbf{Cross-Dataset Validation}: Systematic evaluation on diverse network environments including IoT frameworks \cite{ma2025fsllm}, cloud infrastructures \cite{gaitan2023explainable}, and specialized domains like industrial control systems. This addresses the call in survey literature for validation across heterogeneous environments \cite{buczak2016survey, khraisat2019survey}.

\item \textbf{Real-World Deployment Testing}: Transition from benchmark datasets to real-time streaming environments, addressing the gap between controlled evaluation and operational deployment noted in methodological critiques \cite{arp2020dos}.

\item \textbf{Adaptation to Evolving Threats}: Extension of the framework to handle concept drift and zero-day attacks through continuous learning mechanisms, addressing limitations of static dataset evaluation.

\item \textbf{Domain-Specific Optimization}: Investigation of how strategic sampling, feature selection, and explainability components require adjustment for different network topologies and threat models.

\item \textbf{Integration with Threat Intelligence Feeds}: Enhancement of the framework with external contextual information to improve detection accuracy and relevance in operational settings.
\end{itemize}

This framework represents a promising stepping-stone in the pursuit of trust-worthy, explainable AI within cybersecurity by reconciling strong detection performance on benchmark data with reasonable methodological choices. Combining strategic sampling, automatic leakage prevention and full explainability fills central flaws of current IDS research pipelines. We will need further validation across a variety of environments to verify broad operational readiness, but our method offers a tightly scrutinized path toward functional, explainable security AI systems.

\bibliographystyle{IEEEtran}
\bibliography{references}

\end{document}